\documentclass[conference]{IEEEtran}
\IEEEoverridecommandlockouts
\usepackage{cite}
\usepackage{amsmath,amssymb,amsfonts}
\usepackage{graphicx}

\usepackage{algorithm}
\makeatletter
\g@addto@macro\fs@ruled{%
  \def\@fs@pre{\vspace*{6pt}\hrule height.8pt depth0pt\kern2pt}%
  \def\@fs@post{\kern2pt\hrule\relax\vspace*{-5pt}}%
}
\restylefloat{algorithm}
\makeatother

\usepackage{algorithmic}
\usepackage{textcomp}
\usepackage{xcolor}
\usepackage{tabularx}
\usepackage{booktabs} 
\usepackage{placeins}
\usepackage{stfloats}
\usepackage{cuted}
\usepackage{capt-of}
\usepackage{url}

\def\IEEEtitletopspaceextra{0.25in}

\begin{document}
\title{Optimizing Team Behavior via Extremum-seeking Control in Multi-agent Reinforcement Learning}

\author{    
    \IEEEauthorblockN{Svar Rajankumar Patel}
    \IEEEauthorblockA{
        Department of Computer Science \\
        University of Massachusetts Lowell \\
        Lowell, Massachusetts, USA
    }
    \and
    \IEEEauthorblockN{Kshitij Jerath}
    \IEEEauthorblockA{
        Department of Mechanical and Industrial Engineering \\
        University of Massachusetts Lowell \\
        Lowell, Massachusetts, USA
    }
}

\maketitle

\begin{abstract}
In cooperative Multi-Agent Reinforcement Learning (MARL), agents must learn to coordinate their actions to achieve shared objectives. A key indicator of effective coordination is behavioral diversity, i.e., the agents' potential to enact distinct yet complementary control policies given the same set of observations.
Although behavioral diversity can enable role specialization and richer joint behaviors, its appropriate level is highly context- and task-dependent and often costly to tune. 
We propose Adaptive Diversity Control (ADiCo), a two-timescale framework that adapts behavioral diversity online during training. On a fast timescale, Diversity Control (DiCo) regulates system diversity toward a target value. On a slower timescale, an Extremum Seeking Controller (ESC) adjusts this target using only mean episodic returns, without requiring an explicit model or analytical gradient of the diversity-performance landscape.
We evaluate ADiCo on six configurations across two cooperative MARL tasks, where it maintains task-appropriate diversity levels: low diversity when homogeneous policies are beneficial and higher diversity when heterogeneous policies are required. Through this adaptation, ADiCo matches or exceeds the mean performance of DiCo (best grid-searched diversity target) and Unconstrained DiCo (diversity allowed to emerge without target regulation). These results demonstrate that incorporating classical feedback control into MARL training enables effective coordination by providing a systematic alternative to task-specific manual selection of the diversity target.

\end{abstract}


\section{Introduction}

In cooperative multi-agent reinforcement learning (MARL)~\cite{MARLSurveyComprehensive, MARLSurveyApplication}, a growing body of work has shown that behavioral diversity among agents can provide significant benefits, including the emergence of complementary roles, improved exploration, and greater resilience to disruptions~\cite{RewardSharingRelational,DiCo,WhenDiversity,EffectiveDiversity,UnifiedDiversity}. Under the Centralized Training with Decentralized Execution (CTDE) paradigm, centralized information may be used during training while agents execute decentralized policies based on local observations. The resulting degree of behavioral diversity is shaped by architectural and learning-design choices such as parameter sharing, auxiliary objectives, and intrinsic rewards~\cite{TeamInteractions}. However, these approaches generally induce or encourage diversity without explicitly regulating its resulting magnitude to a desired, measurable level. Consider a team of delivery drones: if all packages are destined for the same neighborhood, similar behaviors may be sufficient, resulting in low diversity. In contrast, if packages must be delivered to different parts of a city, agents may need to specialize their routes, requiring greater behavioral diversity. The appropriate degree of diversity therefore depends on the task, motivating methods that can explicitly measure and regulate behavioral diversity according to task requirements.

In this work, we measure behavioral diversity using System Neural Diversity (SND)~\cite{SND}, which aggregates pairwise differences between agents' policies into a single system-level measure. To regulate this diversity, we use Diversity Control (DiCo)~\cite{DiCo}, which represents each agent's policy as the sum of a shared component and a dynamically scaled agent-specific component, allowing system diversity to be controlled toward a prescribed SND target. However, the appropriate diversity target is task-dependent and generally unknown a priori. We therefore formulate its selection as a feedback optimization problem and use Extremum Seeking Control (ESC)~\cite{ESC_Stability}, a model-free optimization technique used across domains including robotics, manufacturing and energy systems~\cite{ESC_Applications,MulticopterESC}. ESC applies small sinusoidal perturbations to the diversity target and processes the resulting variations in task performance to estimate the local direction of improvement, which is then used to adapt the nominal diversity setpoint.

Building on these components, we propose Adaptive Diversity Control (ADiCo), a two-timescale framework in which DiCo regulates diversity while ESC adapts the diversity target on a slower timescale. ADiCo preserves DiCo's compatibility with on- and off-policy actor-critic algorithms using deterministic or stochastic continuous-action policies~\cite{DiCo}, including IPPO~\cite{IPPO} and MADDPG~\cite{MADDPG}, without modifying the underlying MARL learning objective. The main contributions of this work are as follows:

\begin{itemize}
    \item We introduce ADiCo, a two-timescale framework that integrates ESC with DiCo to adapt behavioral diversity during cooperative MARL training using only scalar performance feedback, without requiring an explicit model or analytical gradient of the diversity--performance relationship.

    \item We validate ADiCo on six configurations across two cooperative MARL tasks, demonstrating task-appropriate diversity adaptation and performance comparable to or better than grid-searched and unconstrained DiCo baselines without manual tuning of the diversity target.
    
\end{itemize}

\section{Background and Related Work}

Behavioral diversity in cooperative MARL has been encouraged through a range of architectural and learning mechanisms. Selective parameter-sharing and agent-specific architectures allow policies to specialize to different degrees~\cite{SePS,SNP_PS,Kaleidoscope,HyperMARL}, while methods based on latent variables, role representations, and intrinsic objectives explicitly encourage diverse behaviors or specialization~\cite{MAVEN,ROMA,SocialInfluence}. These approaches primarily enable or promote heterogeneity, but generally do not expose the resulting magnitude of behavioral diversity as an explicit control variable. System Neural Diversity (SND)~\cite{SND} addresses the measurement problem by quantifying policy-level heterogeneity through pairwise behavioral distances, while Diversity Control (DiCo)~\cite{DiCo} uses this measure to regulate the system toward a prescribed diversity target, $\mathrm{SND}_{\mathrm{des}}$. Although DiCo outlines an outer-loop procedure for adjusting desired diversity toward a user-specified performance level, it does not address the problem of locating an unknown reward-maximizing diversity target. This motivates adapting the desired diversity directly from task-performance feedback during training.

Adaptive hyperparameter optimization has also been studied within reinforcement learning. Population-Based Training (PBT) jointly adapts model parameters and hyperparameter schedules using a population of concurrently trained agents~\cite{PBT}, while Hyperparameter Optimization on the Fly (HOOF) adapts policy-gradient hyperparameters within a single training run by evaluating candidate policy updates using off-policy estimates~\cite{HOOF}. Meta-gradient methods instead differentiate through the learning process to adapt differentiable hyperparameters online~\cite{MetaGradientRL}. In our setting, the desired diversity is a control input to DiCo, and its effect on task performance is unknown and may change as policies learn. We therefore require a model-free controller that can adjust this target during training using scalar performance feedback, without access to an analytical gradient of the diversity-performance relationship.

Extremum Seeking Control (ESC) meets these requirements by applying small periodic perturbations and using the resulting performance variations to estimate a local direction of improvement~\cite{ESC_Stability,ESC_Applications}. ESC has previously been applied to dynamical systems~\cite{MulticopterESC} and to refining sampled actions during reinforcement learning~\cite{ChangGaoESC}. In ADiCo, ESC instead adapts a system-level diversity target on a slower timescale than the underlying DiCo/MARL process, seeking a locally high-performing diversity level along a single training run.

\section{Adaptive Diversity Control}
In this section, we present Adaptive Diversity Control (ADiCo), a framework for adapting behavioral diversity during cooperative MARL training. 

\begin{figure*}[t]
    \centering
    \includegraphics[width=0.85\textwidth]{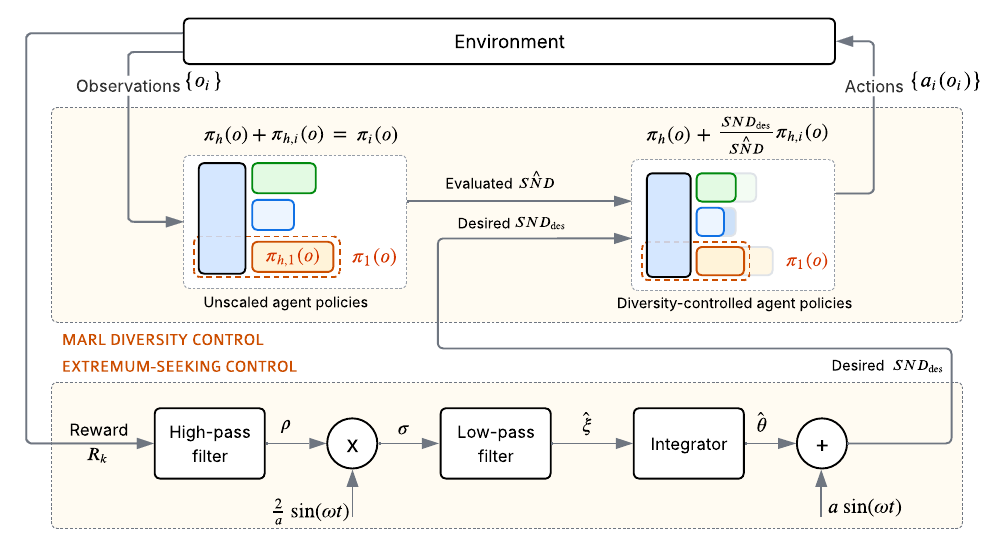}
    \caption{ADiCo system architecture. The DiCo policy decomposition and rescaling pipeline (top) is retained from~\cite{DiCo}, where policies are rescaled to match the desired diversity $\mathrm{SND}_{\mathrm{des}}$, with the current diversity estimated via exponential moving average. Note the illustrative rescaling of Agent 1's policy $\pi_1(o)$ in the pipeline (top). The ESC signal chain (bottom) processes the mean reward to estimate the gradient and update the diversity setpoint. The complete procedure is described in Algorithm~\ref{alg:ad2c}.}
    \label{fig:framework}
\end{figure*}

\subsection{Problem Formulation}
We model the cooperative MARL problem as a Decentralized Partially Observable Markov Decision Process (Dec-POMDP), defined by the tuple $\langle \mathcal N,S,\{A_i\},\{O_i\},P,R,\gamma\rangle$, where $\mathcal {N} = \{1,\ldots,N\}$ denotes the set of agents, $S$ is the global state space, $A_i$ and $O_i$ denote the action and observation spaces associated with agent $i$, respectively. $P$ is the transition function, $R$ is the shared team reward, and $\gamma\in[0,1)$ is the discount factor. At each timestep $t$, agent $i$ receives a local observation $o_{i,t}\in O_i$ and selects an action $a_{i,t}\in A_i$ according to its decentralized policy $\pi_i(a_{i,t}\mid o_{i,t})$. The joint policy $\boldsymbol{\pi}=(\pi_1,\ldots,\pi_N)$ is trained to maximize the expected discounted team return $\max_{\boldsymbol{\pi}}\mathbb{E}_{\boldsymbol{\pi}}\left[\sum_{t=0}^{\infty}\gamma^t R(s_t,\mathbf{a}_t)\right]$, where $\mathbf{a}_t=(a_{1,t},\ldots,a_{N,t})$ denotes the joint action at environment timestep $t$. Training follows the Centralized Training with Decentralized Execution (CTDE) paradigm, in which centralized information may be used during training while each agent executes its policy using only local observations.

\subsection{Measuring and Controlling Behavioral Diversity}

To quantify behavioral diversity, we employ System Neural Diversity (SND)~\cite{SND}, defined as the average pairwise behavioral distance between agents over a shared observation set $O$. The behavioral distance between two agents is measured using the 2-Wasserstein distance $W_2$. For the stochastic Gaussian policies considered in this work, this distance has the closed-form expression

\begin{equation}
W_2(\pi_i(o), \pi_j(o)) = \sqrt{ \|\mu_i(o)-\mu_j(o)\|_2^2 + \|\sigma_i(o)-\sigma_j(o)\|_2^2 },
\label{eq:wasserstein}
\end{equation}

\noindent where $\mu_i(o)$ and $\sigma_i(o)$ denote the mean and standard deviation vectors of agent $i$'s action distribution at observation $o$. For deterministic policies, the variance term vanishes and $W_2$ reduces to the Euclidean distance between the agents' policy outputs. The system-level diversity is obtained by averaging these pairwise distances across all agent pairs and observations:

\begin{equation}
\mathrm{SND}(\{\pi_i\}_{i\in\mathcal N}) = \frac{2}{N(N-1)|O|} \sum_{i<j}\sum_{o\in O} W_2(\pi_i(o),\pi_j(o)).
\label{eq:snd}
\end{equation}

\noindent To control this diversity, DiCo~\cite{DiCo} decomposes each agent's policy into a shared homogeneous component $\pi_h(o)$ and an agent-specific heterogeneous component $\pi_h^i(o)$. We denote the desired diversity target $\mathrm{SND}_{\mathrm{des}}$ by $\theta$. DiCo dynamically rescales the heterogeneous component according to (Eq.~\eqref{eq:dico_rescaling}):

\begin{equation}
\pi_i(o) = \pi_h(o) + \frac{\theta}{\widehat{\mathrm{SND}}} \pi_h^i(o),
\label{eq:dico_rescaling}
\end{equation}

\noindent where $\widehat{\mathrm{SND}}$ is a smoothed estimate of the diversity of the unscaled heterogeneous components $\{\pi_h^i\}_{i\in\mathcal N}$, computed from each training batch and updated at every policy update using the DiCo soft-update mechanism~\cite{DiCo}. The scaling factor $\theta/\widehat{\mathrm{SND}}$ adjusts the contribution of the agent-specific component so that the realized system diversity is regulated toward the desired level $\theta$. Thus, DiCo reduces behavioral-diversity regulation to a single scalar target, while the task-appropriate value of that target remains to be determined.

\subsection{Diversity-Performance Landscape}

Although DiCo regulates behavioral diversity toward a prescribed target $\theta$, the target that maximizes team performance is generally unknown and task-dependent. To formulate its selection, we introduce the diversity--performance map $J(\theta)$, representing the expected team performance associated with a diversity target $\theta$. This relationship may change as the policies learn. We therefore consider it on the slower ESC timescale, with the underlying MARL/DiCo training process treated as the faster subsystem.

At each cycle $k$, the diversity target is held fixed while agents collect experience and update their policies. The updated policies are then held fixed and evaluated over $M$ complete episodes to obtain the mean episodic return $R_k$. This return provides a stochastic performance measurement at the current target, which ESC uses to adjust the target for the next cycle. Thus, diversity adaptation occurs after each training and evaluation cycle, rather than at every environment timestep.

Using this feedback, ADiCo seeks a locally optimal diversity target $\theta^\star$ within the admissible range $\Theta=[0,\theta_{\max}]$. Because the diversity--performance map is unknown and its analytical gradient is unavailable, we treat target selection as a black-box optimization problem. ESC addresses this problem by perturbing the target and processing the resulting changes in $R_k$ to estimate a local direction of performance improvement.

\subsection{Extremum Seeking Diversity Adaptation}
Having formulated diversity-target selection as a black-box optimization problem, ADiCo uses ESC to estimate a local direction of performance improvement through periodic perturbations. At each ESC cycle $k$, a sinusoidal perturbation with amplitude $a$ and phase $\phi_k$ is applied to the nominal setpoint $\hat{\theta}_k$. The perturbed value is projected onto $\Theta$ to obtain the diversity target $\theta_k$ supplied to DiCo. The phase advances according to $\phi_{k+1}=\phi_k+\omega\Delta t$, where $\omega$ is the perturbation angular frequency and $\Delta t$ is the interval between ESC updates.

After training and evaluation at $\theta_k$, the measured return $R_k$ is processed to estimate the local performance response to the perturbation. A high-pass filter (HPF) attenuates slowly varying components of the return, demodulation extracts the response correlated with the perturbation, and a low-pass filter (LPF) attenuates the remaining oscillatory components. The resulting signal $\hat{\xi}_k$ estimates the local direction of performance improvement.

ESC uses $\hat{\xi}_k$ to update the nominal setpoint $\hat{\theta}_k$, with a positive integrator gain $\eta$ controlling the adaptation rate and projection onto $\Theta$ enforcing the admissible range. These updates seek a locally high-performing diversity level. The complete procedure is summarized in Algorithm~\ref{alg:ad2c}.

\newcommand{\phase}[1]{%
    \item[]\textcolor{teal}{\textbf{#1}}%
}

\begin{algorithm}[t]
\caption{Adaptive Diversity Control (ADiCo)}
\label{alg:ad2c}
\begin{algorithmic}[1]

\REQUIRE Dither amplitude $a$, perturbation frequency $\omega$,
ESC update interval $\Delta t$, integrator gain $\eta$,
filter coefficients $\alpha_h,\alpha_l$, evaluation episodes $M$,
initial diversity setpoint $\hat{\theta}_0$,
admissible range $\Theta=[0,\theta_{\max}]$

\STATE \textbf{Initialize:}
$\phi_0=0$, $R_{-1}=0$, $\rho_{-1}=0$, $\hat{\xi}_{-1}=0$

\FOR{ESC cycle $k=0,1,\ldots,K-1$}

\phase{Phase 1: Set Diversity Target}

\STATE \textit{\textbf{Apply dither modulation and enforce admissibility constraint:}}
\STATE \hspace{\algorithmicindent}
$\theta_k \leftarrow
\Pi_{\Theta}\!\left(
\hat{\theta}_k+a\sin(\phi_k)
\right)$

\phase{Phase 2: DiCo Policy Training}

\FOR{each MARL training iteration}
    \STATE Collect rollouts and update actor--critic parameters
    \STATE \textit{\textbf{Regulate diversity toward $\theta_k$:}}
    \STATE \hspace{\algorithmicindent}
    $\pi_i(o)\leftarrow
    \pi_h(o)+
    \dfrac{\theta_k}{\widehat{\mathrm{SND}}}\pi_h^i(o)$
\ENDFOR

\phase{Phase 3: ESC Setpoint Update}

\STATE \textit{\textbf{Evaluate policy performance:}}
\STATE \hspace{\algorithmicindent}
Evaluate policies over $M$ episodes and compute $R_k$

\STATE \textit{\textbf{Estimate performance gradient:}}

\STATE \hspace{\algorithmicindent}
\makebox[0.63\linewidth][l]{$
\rho_k \leftarrow
\alpha_h\bigl(\rho_{k-1}+R_k-R_{k-1}\bigr)
$}
$\triangleright$ HPF

\STATE \hspace{\algorithmicindent}
\makebox[0.63\linewidth][l]{$
\sigma_k \leftarrow
\dfrac{2}{a}\rho_k\sin(\phi_k)
$}
$\triangleright$ Demodulation

\STATE \hspace{\algorithmicindent}
\makebox[0.63\linewidth][l]{$
\hat{\xi}_k \leftarrow
\alpha_l\sigma_k+(1-\alpha_l)\hat{\xi}_{k-1}
$}
$\triangleright$ LPF

\STATE \textit{\textbf{Update nominal setpoint:}}
\STATE \hspace{\algorithmicindent}
$\hat{\theta}_{k+1}
\leftarrow
\Pi_{\Theta}\!\left(
\hat{\theta}_k+\eta\hat{\xi}_k
\right)$

\STATE \hspace{\algorithmicindent}
$\phi_{k+1}\leftarrow\phi_k+\omega\Delta t$

\ENDFOR
\end{algorithmic}
\end{algorithm}

\vspace*{-6pt}

\section{Experiments}

This section evaluates ADiCo's ability to adapt behavioral diversity across different task configurations and analyzes its convergence behavior.

\subsection{Environments and Task Configurations}

\begin{figure}[ht]
\centering
\includegraphics[width=1.0\columnwidth, keepaspectratio]{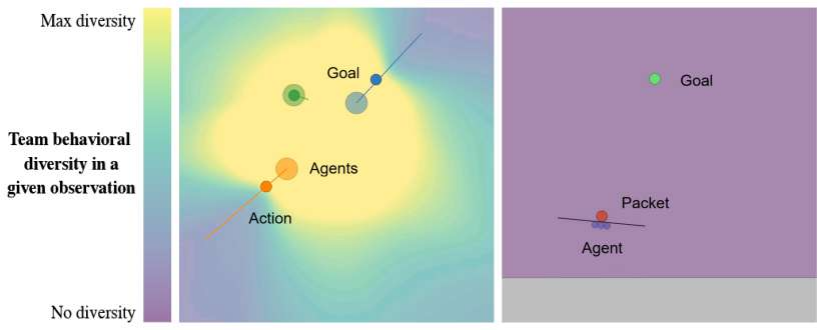}
\caption{\textbf{Navigation} (left) and \textbf{Balance} (right). In Navigation, agents (colored circles) apply continuous 2D forces to reach goal locations (smaller circles). In Balance, agents (blue) coordinate forces on a shared platform (grey) to move the packet (red) toward the goal (green). The background shows behavioral diversity measured using SND~\cite{SND}, with yellow indicating high and purple indicating low diversity.}
\label{fig:task_environments}
\end{figure}

We evaluate ADiCo on two cooperative continuous-control tasks from the Vectorized Multi-Agent Simulator (VMAS)~\cite{VMAS}: Navigation and Balance (Fig.~\ref{fig:task_environments}). All experiments use three agents under partial observability, with episodes lasting at most 100 steps. We keep the team size fixed to simplify the analysis, although ADiCo is not limited to a fixed number of agents.

For each configuration, we compare ADiCo with two baselines. \textit{Grid-searched DiCo} uses the best fixed SND target from $\{0,0.3,0.5,0.8,1.0,1.3,1.5,1.8,2.0\}$, with a separate 12M-frame training run for each target. \textit{Unconstrained DiCo} fixes the agent-specific scaling factor at one, allowing diversity to emerge without an explicit SND target. These baselines test whether online adaptation can match or improve upon both the best grid-searched target and freely emerging diversity. All methods use the same MARL algorithm, network architecture, and 12M-frame budget per training run.

\begin{figure*}[ht]
\centering
\includegraphics[width=\textwidth, height=7.9cm]{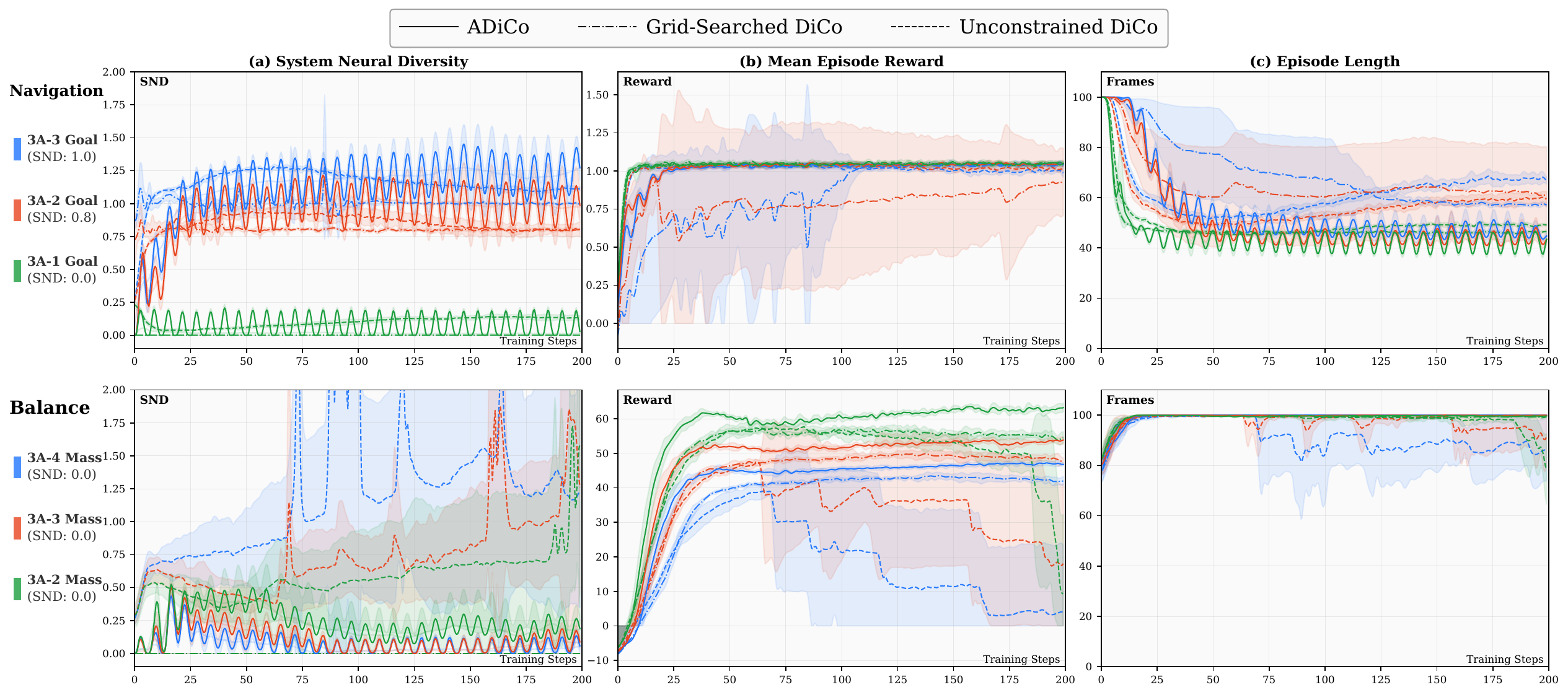}
\caption{Training curves across Navigation (top) and Balance (bottom) tasks comparing ADiCo (solid), Grid-Searched DiCo (dash-dot), and Unconstrained DiCo (dashed). Colors denote task configurations; the left panel shows DiCo's manually tuned SND target for each. (a) System Neural Diversity. (b) Mean episode reward. (c) Episode length.}
\label{fig:main_results}
\vspace{-15pt}
\end{figure*}

\subsection{Training and Evaluation}
All our approaches use IPPO~\cite{IPPO,MAPPO} and are trained for 12M frames over five random seeds. Training is divided into cycles, with each cycle collecting 60,000 frames from 600 parallel environment instances followed by the MARL policy update. The updated policies are then evaluated over $M=200$ episodes, and the mean episodic return $R_k$ is used by ADiCo to update the diversity setpoint before the next cycle.

\section{Results and Analysis}
In this section, we evaluate ADiCo across three key aspects: performance across task configurations, diversity adaptation from different initial targets, and training behavior under continuous ESC perturbations.

\subsection{Performance Across Task Configurations}

We evaluate ADiCo across six configurations of two cooperative tasks. In Navigation, we vary the number of goals (1, 2, or 3) to change the need for specialization, using 3A-$x$G to denote three agents with $x$ goals. In Balance, we vary ball Mass (2, 3, or 4) to change task difficulty while keeping the coordination structure the same, using 3A-$x$M to denote three agents with ball Mass $x$. In Navigation (Fig.~\ref{fig:main_results}, top row), ADiCo adapts to high diversity (SND $\approx$1.2--1.3) for 3A-3G, moderate diversity (SND $\approx$0.8) for 3A-2G, and near-zero diversity for 3A-1G. It achieves reward comparable to grid-searched DiCo, with lower reward variance than Unconstrained DiCo. In Balance (bottom row), ADiCo maintains low diversity across all Mass configurations and matches the performance of DiCo with SND = 0. For Unconstrained DiCo, increasing diversity accompanies declining reward, with performance collapsing in the 3A-4M configuration. These results show that ADiCo automatically adapts toward low diversity when similar behaviors are beneficial and higher diversity when specialization helps. It achieves performance comparable to manually tuned DiCo without requiring configuration-specific grid search, while avoiding the excessive diversity observed with Unconstrained DiCo.

\subsection{Convergence From Different Initializations}

\begin{figure}[ht]
\centering
\includegraphics[width=1.0\columnwidth]{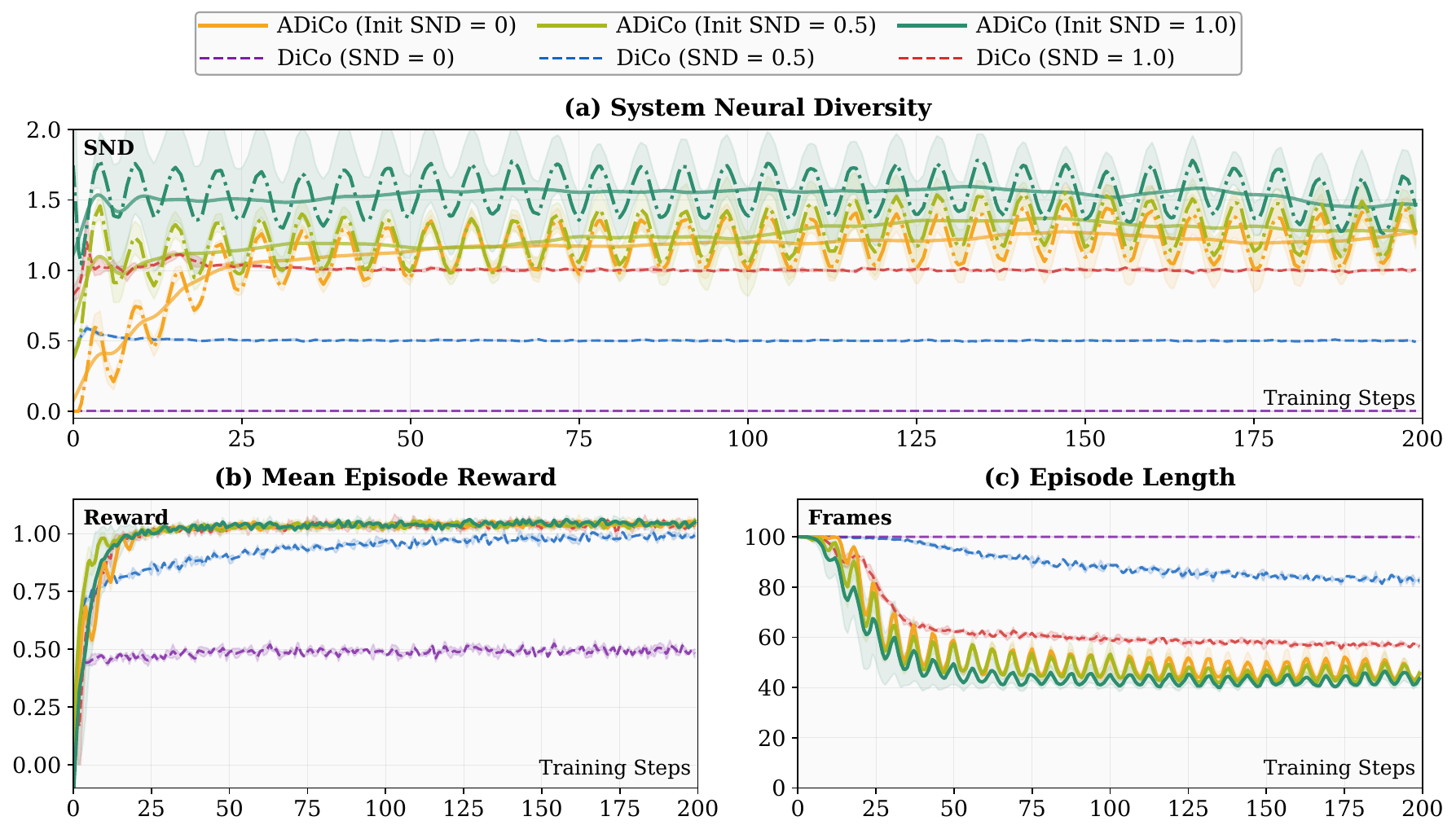}
\caption{Convergence analysis on Navigation 3A-3G comparing ADiCo (solid lines) against DiCo (dashed lines) at SND levels 0, 0.5, and 1.0. (a) SND trajectories. (b) Mean episode reward. (c) Episode length.}
\label{fig:convergence}
\vspace{-10pt}
\end{figure}

The previous experiments initialize ADiCo at SND = 0. To examine whether this choice affects the final diversity level, we test initial targets of 0, 0.5, and 1.0 on Navigation 3A-3G and compare them with DiCo using the same fixed targets. Fig.~\ref{fig:convergence}(a) shows that ADiCo reaches a similar SND range ($\approx$1.2--1.3) from all three initial targets. Runs initialized at 0.5 and 1.0 reach this range faster than those initialized at 0. The DiCo baselines remain near their fixed targets throughout training. All three ADiCo initializations achieve mean returns near 1.0 and episode lengths of approximately 45--50 steps (Fig.~\ref{fig:convergence}(b,c)). DiCo with SND = 0 or 0.5 achieves lower performance, while the best grid-searched target, SND = 1.0, produces longer episodes than ADiCo. These results show that ADiCo reaches similar diversity levels and performance across the tested initializations, while completing episodes faster than the fixed-target baselines. We next examine whether keeping ESC active remains beneficial after diversity settles.

\subsection{Effect of Continued ESC Adaptation}

\begin{figure}[ht]
\centering
\includegraphics[width=1.0\columnwidth, keepaspectratio]{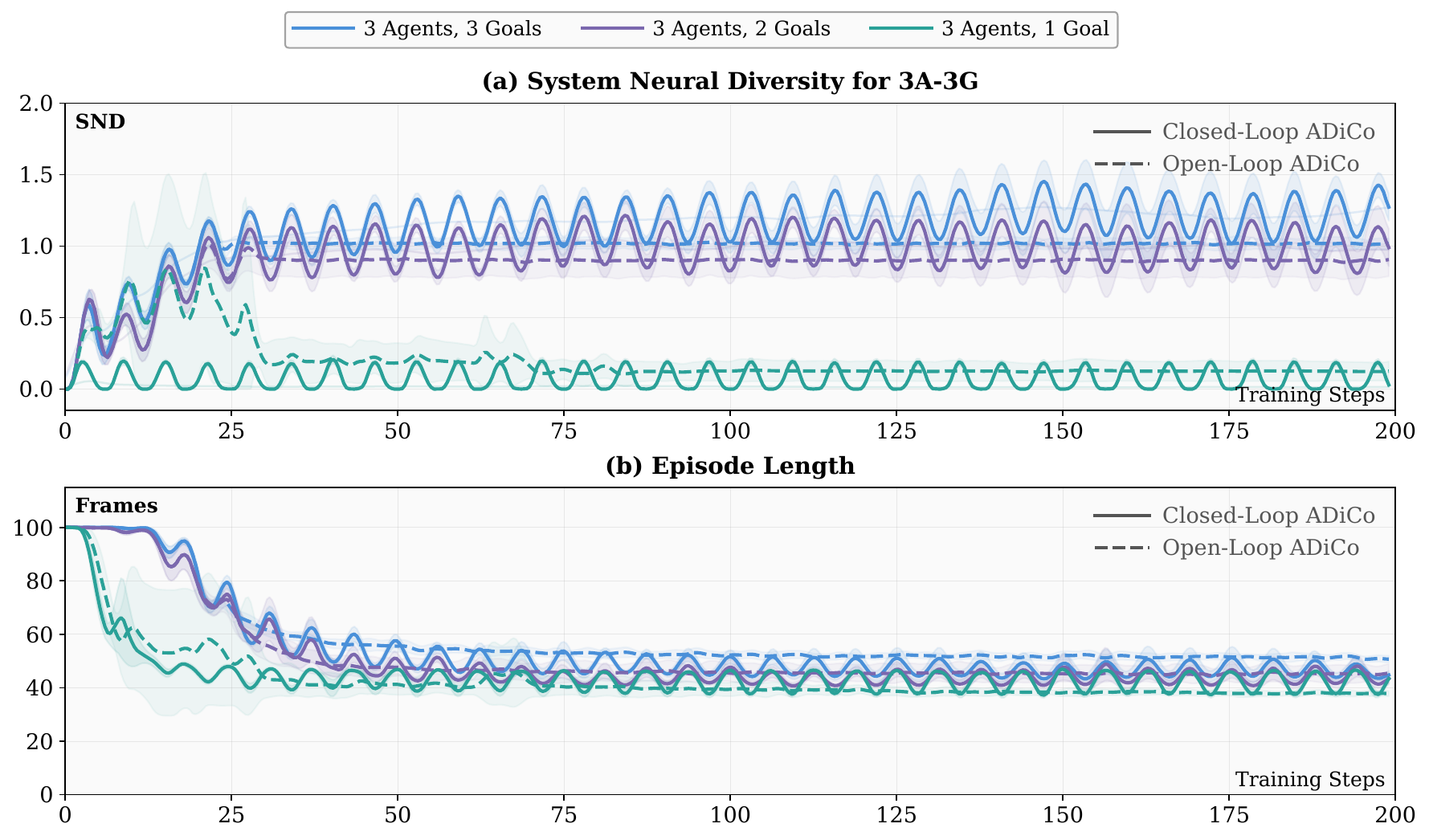}
\caption{Comparison across three Navigation configurations between ADiCo with continued ESC adaptation (Closed-Loop ADiCo) and a variant that disables perturbations and freezes the diversity target when its stopping criterion is met (Open-Loop ADiCo). (a) SND trajectories. (b) Episode length.}
\label{fig:stability}
\end{figure}

ESC continues to perturb the diversity target even after it settles near a particular level. To examine whether continued adaptation remains useful, we compare ADiCo with a variant that monitors the gradient estimate and disables perturbations when a stopping criterion is met. This variant then holds the diversity target fixed for the remainder of training. Fig.~\ref{fig:stability}(a) shows that the frozen-target variant maintains SND near zero for 3A-1G, approximately 0.9 for 3A-2G, and approximately 1.0 for 3A-3G. These values are close to the best targets found through grid search. Once the target is frozen, the SND trajectories flatten as expected. In contrast, ADiCo with continued ESC adaptation achieves shorter episodes in 3A-3G and comparable episode lengths in the other configurations (Fig.~\ref{fig:stability}(b)). These results suggest that continued adaptation can remain beneficial after diversity appears to settle. However, this comparison does not separate the effect of perturbations from the benefit of further target updates, since the frozen-target variant disables both.

\section{Limitations and Future Directions}

While ADiCo uses mean episodic return to adapt diversity, changes in return are not always caused by diversity adjustments alone. Factors such as policy learning, environmental shifts, and adapting opponents also influence performance, complicating adaptation in adversarial or mixed-motive domains. Consequently, future work could explore alternative feedback signals and evaluation methods that better isolate the specific effect of diversity. Furthermore, because ADiCo relies on extremum seeking control, it risks converging to a local optimum of the performance signal rather than the globally optimal diversity target for the task. This limitation is compounded by the fact that ADiCo currently tunes only a single team-wide SND target, whereas distinct agents or roles may benefit from different diversity levels. Extending the framework with multidimensional ESC could allow multiple diversity targets to be adjusted from a shared performance signal, enabling more detailed control over team behavior. Finally, ADiCo adapts diversity exclusively during training rather than runtime execution. Because teams often need to alter behavior as conditions change within an episode, integrating ADiCo with event-based policy adaptation~\cite{EventDiversity} could allow agents to dynamically adjust their diversity in response to runtime events.

\section{Concluding Remarks}

In this work, we introduced ADiCo, a framework that adapts team diversity during training using team-performance feedback. Across two cooperative tasks, ADiCo adjusts diversity to task-appropriate levels. Experiments with different initial setpoints further show that, in the tested configuration, users can begin with a rough diversity estimate and let ADiCo refine it without an explicit model of team dynamics. More broadly, ADiCo illustrates how model training can be viewed as a process whose adjustable components are optimized through feedback, much like machines in industrial systems. This perspective allows us to draw on established control techniques to improve the adaptability of learning systems.

\end{document}